\documentclass[floats,floatfix,showpacs,amssymb,prd,superscriptaddress,nofootinbib,twocolumn,aps]{revtex4-1}
\usepackage{graphicx, epsfig, amssymb} 
\usepackage{amsmath, amsfonts}
\usepackage{bm} 

\usepackage[linktocpage]{hyperref}
\usepackage[caption=false]{subfig}
\usepackage[usenames]{color}

\def\be{\begin{equation}}
\def\ee{\end{equation}}
\def\beq{\begin{eqnarray}}
\def\eeq{\end{eqnarray}}

\newcommand{\bea}{\begin{eqnarray}}
\newcommand{\eea}{\end{eqnarray}}
\newcommand{\ben}{\begin{enumerate}}
\newcommand{\een}{\end{enumerate}}
\newcommand{\bi}{\begin{itemize}}
\newcommand{\ei}{\end{itemize}}

\newcommand{\nn}{\nonumber}

\begin{document}

\title{\large Compact stars in Eddington inspired gravity}

 \author{Paolo Pani}
 \affiliation{CENTRA, Departamento de F\'{\i}sica, 
 Instituto Superior T\'ecnico, Universidade T\'ecnica de Lisboa - UTL,
 Av.~Rovisco Pais 1, 1049 Lisboa, Portugal.}
 \author{Vitor Cardoso}
 \affiliation{CENTRA, Departamento de F\'{\i}sica, 
 Instituto Superior T\'ecnico, Universidade T\'ecnica de Lisboa - UTL,
 Av.~Rovisco Pais 1, 1049 Lisboa, Portugal.}
 \affiliation{Department of Physics and Astronomy, The University of Mississippi, University, MS 38677, USA.}

 \author{T\'erence Delsate}
 \affiliation{CENTRA, Departamento de F\'{\i}sica, 
 Instituto Superior T\'ecnico, Universidade T\'ecnica de Lisboa - UTL,
 Av.~Rovisco Pais 1, 1049 Lisboa, Portugal.}


\begin{abstract}
A new, Eddington inspired theory of gravity was recently proposed by Ba\~nados and Ferreira. It is equivalent to General Relativity in vacuum, but differs from it inside matter. This viable, one parameter theory was shown to avoid cosmological singularities and turns out to lead to many other exciting new features that we report here. First, for a positive coupling parameter, the field equations have a dramatic impact on the collapse of dust, and do {\it not} lead to singularities. We further find that the theory supports stable, compact pressureless stars made of perfect fluid, which provide interesting models of self-gravitating dark matter. Finally, we show that the mere existence of relativistic stars imposes a strong, near optimal constraint on the coupling parameter, which can even be improved by observations of the moment of inertia of the double pulsar.
\end{abstract}


\maketitle
\date{today}
\noindent{\bf{\em I. Introduction.}}
Einstein's General Relativity (GR) is able to explain a wide variety of phenomena at solar system scales and beyond,
and after decades of intense scrutiny stands as the most attractive theory of gravity. 
However, high-curvature corrections may be necessary to address unresolved issues, such as the presence of singularities in cosmology and in the interior of black holes.

Recently, an intriguing alternative to GR (based on an original proposal by Eddington) was put forward by Ba\~nados and Ferreira (BF)~\cite{Banados:2010ix} (see also~\cite{Vollick:2005gc}). BF theory is described by the action
\be \label{action}
S=\frac{2}{\kappa}\int d^4x\,\left(\sqrt{-\left|g_{ab}+\kappa R_{ab}(\Gamma)\right|}-\lambda\sqrt{-g}\right)\,,
\ee
where $R_{ab}(\Gamma)$ denotes the symmetric part of the Ricci tensor, built from the connection $\Gamma_{ab}^c$ and $\lambda$ is related to the cosmological constant, $\Lambda=(\lambda-1)/\kappa$. We will focus on asymptotically flat solutions and set $\lambda=1$.
The metric $g$ and the connection $\Gamma$ are independent fields and, at the classical level, matter is minimally coupled to the metric only. The BF proposal explores the fact that the coupling between matter and gravity is one of the least tested sectors of gravitation;
in fact the theory can be shown to be {\it completely equivalent to GR in vacuum}~\cite{Banados:2010ix}. 
However, it dramatically differs from GR in the presence of matter, for example it yields a singularity-free cosmology, thus presenting itself as a potentially exciting gravity theory. The BF theory modifies the Newtonian regime (again, in the matter coupling), but tests of gravity within matter are extremely hard to carry, partly because we understand the coupling to matter so poorly. Thus, constraints on the BF theory as a result of Earth-based experiments are hard to accomplish. 

Here we show that non-linear effects and deviations from Einstein's theory are more pronounced
inside high-density objects. When $\kappa R_{ab}\ll1$, the first corrections to the Einstein equations read
\begin{equation}
R_{ab}(\Gamma)=T_{ab}-\frac{1}{2}Tg_{ab}+\kappa\left[S_{ab}-\frac{1}{4}Sg_{ab}\right]+{\cal O}(\kappa^2)\,,\nn
\end{equation}
where two types of ${\cal O}(\kappa)$ corrections appear: those hidden in $R_{ab}(\Gamma)$, which implicitly depend on derivatives of matter fields, and those depending on $S_{ab}={T^c}_a T_{c b}-\frac{1}{2}T T_{ab}$, which are quadratic in the matter fields. 
Hence we expect strong corrections at high densities or where strong matter gradients exist, for example in early cosmology~\cite{Banados:2010ix} or inside neutron stars (NSs).
Higher order corrections in the matter fields were also discussed in Ref.~\cite{Kerner:1982} to cure cosmological singularities.
The purpose of this letter is to show that the best possible constraints on the theory arise from the study of NSs and other compact objects.
In the process, we report new remarkable features of BF theory. 

\noindent{\bf{\em IIA. Stars in the non-relativistic limit.}}
Let us start by discussing the nonrelativistic limit of~\eqref{action}. The modified Poisson equation reads~\cite{Banados:2010ix}
\be
\nabla^2\Phi=4\pi G\rho+\kappa\,\nabla^2\rho/4\,.\label{Poisson}
\ee
From Eq.~\eqref{Poisson}, and requiring spherical symmetry, the hydrostatic equilibrium equation follows
\be
dP/dr=-Gm(r)\rho/r^2-\kappa\rho\rho'/4\,.\label{hydroeq2}
\ee
While BF corrections are absent for constant density profiles, interesting effects may show up for non-trivial matter distributions.
Newtonian stellar models are solutions of Eq.~\eqref{hydroeq2} supplemented by the standard mass conservation, $dm/dr=4\pi r^2\rho(r)$ and an equation of state (EOS). We note that constant density stars in BF theory are potentially pathological, since they introduce a Dirac delta contribution in Eq.~\eqref{hydroeq2}. For this reason, in this letter we shall focus on more realistic, polytropic models of the form $P(\rho)=K\rho^{(n+1)/n}$,
where $K$ and $n$ are constants.

\noindent{\em IIA1. Newtonian pressureless stars.}
Remarkably, this theory supports pressureless stars, i.e. stars made of non-interacting particles, which provide interesting models for self-gravitating dark matter. Indeed, if $P\equiv0$ and $\kappa>0$, Eq.~\eqref{hydroeq2} is solved by
\be
\rho(r)={\rho_c}\sin(\varpi r)/(\varpi r)\,,\qquad \varpi=4\sqrt{G\pi/\kappa}\,.\label{pressureless_Newtonian}
\ee
The radius and mass of the star read $R=\pi/\varpi$ and $M=4\pi^2\rho_c/\varpi^3$, respectively. 
In the interior, the Newtonian potential is constant and it matches continuously the vacuum potential $M/r$ at the radius. 
Below, we prove that these solutions are also linearly stable.

\noindent{\em IIA2. Newtonian polytropic models.}
For a generic polytropic index $n$, the field equation must be solved numerically, imposing $\rho\sim\rho_c+\rho_2 r^2$ at the center. It is easy to show that realistic stellar configurations (with $\rho\to 0$ at the surface of the star) can only exist provided the following condition is satisfied
\be
\kappa>-|\kappa_c|=-4 K (1+1/n) \rho_c^{-1+1/n}\,.\label{kcrit}
\ee
Similar constraints exist for any EOS for which the pressure increases monotonically with the density.
For $\kappa>0$, condition (\ref{kcrit}) is always fulfilled. 
In some cases the Lane-Emden equation obtained from Eq.~\eqref{hydroeq2} can be solved analytically~\cite{Chandra_Book_Stars}. For instance
if $n=1$, $P(\rho)=K\rho^2$, and the solution reads as in~\eqref{pressureless_Newtonian}, but with $\varpi=4\sqrt{G\pi/(8K+\kappa)}$, so that
it exists for $\kappa>-8K$ and reduces to the pressureless case for $K=0$.

\noindent{\bf{\em IIB. Stability in the non-relativistic limit.}}
We now discuss stability of the Newtonian configurations against {\it radial} perturbations.
The standard treatment can be extended straightforwardly to encompass BF theory~\cite{Shapiro:1983du}.
Assuming a time dependence $\sim e^{i\omega t}$ for the fields, the modified eigenvalue equation reads
\be
\!\!\frac{4\xi P'}{\,r} \!+\!\frac{\kappa\rho}{4}\left[\frac{2}{r}\xi\rho'\!\!-\!\!\xi'\rho'\!\!-\!\!\left[\frac{\rho}{r^2}(r^2\xi)'\right]'\right]\!\!-\!\!\left[\frac{\gamma P}{r^2}(r^2\xi)'\right]'\!\!=\rho\xi\omega^2,\nonumber
\ee
where $\gamma$ is the adiabatic index of the perturbations. This equation
must be solved for the Lagrangian displacement $\xi$ requiring regularity at the center and at the radius.
An instability corresponds to an eigenmode with $\omega^2<0$.

\noindent{\em IIB1. Pressureless stars.}
For $P\equiv0$ and $\rho$ given in~\eqref{pressureless_Newtonian},
our numerical study found eigenmodes with $\omega^2>0$ and no unstable mode. 
Therefore, pressureless stars in the modified Newtonian theory are stable. 
As we shall see, these solutions persist in the fully relativistic theory.

\noindent{\em IIB2. Newtonian polytropic stars.}
In Newtonian gravity, polytropic models with $\gamma=4/3$ are marginally stable for any polytropic index $n$~\cite{Shapiro:1983du}. In our case, these models are stable if $\kappa>0$ and unstable if $\kappa<0$. For generic values of $\gamma$, positive values of $\kappa$ contribute to stabilize the models, while negative values work in the opposite direction.

\begin{figure}[htb]
\begin{center}
\begin{tabular}{c}
\epsfig{file=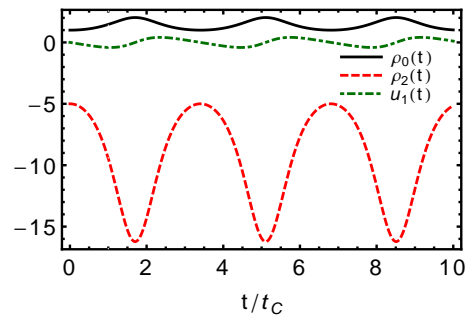,width=6.5cm,angle=0}
\end{tabular}
\caption{Oscillatory behavior of collapse-related quantities as functions of time for $\kappa|\eta|=6.5$. 
\label{fig:collapse}}
\end{center}
\end{figure}
%
\noindent{\bf {\em IIC. Gravitational collapse.}}
The collapse of incoherent dust in the Newtonian limit shares many properties with its relativistic analogue~\cite{Oppenheimer:1939,Florides1977138}. 
The relevant Eulerian equations governing the fluid dynamics are Eqs.~(9) and (11) in Ref.~\cite{Florides1977138} together with
\begin{equation}
 \partial_tu(\mathbf{x})+u(\mathbf{x}) \partial_r u(\mathbf{x})=-{GM(\mathbf{x})}/{r^2}-{\kappa}\partial_r\rho(\mathbf{x})/4\,, \nn
\end{equation}
where $u(\mathbf{x})$ is the fluid velocity and $\mathbf{x}=(t,r)$. In standard Newtonian gravity, $\kappa=0$, the equations can be solved analytically when $\rho(\mathbf{x})=\rho(t)$ and they correspond to the relativistic Oppenheimer-Snyder solution~\cite{Florides1977138}. The dust collapses in a finite time $t_C=\pi\sqrt{R^3/(8M_T)}$, where $R$ and $M_T$ are the initial radius and the total mass of the spherical dust configuration. The density $\rho$ and the fluid velocity $u$ diverge \emph{at any radius} when $t\to t_C$. Hence, we can simply solve for an expansion close to the center
\begin{eqnarray}
 \rho(\mathbf{x})&=&\rho_0(t)+\rho_1(t)r+\rho_2(t) r^2+{\cal O}(r^3)\,,\\
 u(\mathbf{x})&=&u_0(t)+u_1(t)r+u_2(t)r^2+{\cal O}(r^3)\,,
\end{eqnarray}
and the collapse occurs if these fields diverge at some $t$.
The field equations impose $\rho_1(t)=u_0(t)=u_2(t)=0$ and $\rho_2(t)=\eta\rho_0^{5/3}(t)$, where $\eta<0$ is a constant, and
\begin{equation}
 t(\rho_0)-t(\rho_i)=\int_{\rho_{i}}^{\rho_0}\frac{dx x^{-\frac{4}{3}}}{\sqrt{24\pi G}}\sqrt{x^{\frac{1}{3}}-\rho_{i}^{\frac{1}{3}}+\frac{\kappa\eta}{8\pi G}\left(x-\rho_{i}\right)}\,,\nn
\end{equation}
which, for $\kappa=0$, reduces to that in Ref.~\cite{Florides1977138}. This equation can be integrated analytically for any $\kappa$.
A collapse occurs when $t(\rho_0\to\infty)\geq t(\rho_{i})$, i.e. the time corresponding to an infinite density is in the future.  
We find that this condition is fulfilled only when $\kappa\leq0$ but, when $\kappa>0$, the collapse \emph{does not} occur. Fig.~\ref{fig:collapse} shows that in this case the matter fields have an oscillatory behavior, whose period and amplitude depend on $\kappa\eta$. The same feature is found in early cosmology~\cite{Banados:2010ix}. This suggests that singularities may be avoided in BF theory with $\kappa>0$, due to ``repulsive gravity'' effects proportional to $\kappa\rho'$.

\noindent{\bf{\em IIIA. Relativistic compact stars.}}
Let us now consider static and spherically symmetric perfect fluid stars in the fully relativistic theory, described by
\bea
q_{ab}dx^a dx^b &=& -p(r) dt^2 + h(r) dr^2 + r^2 d\Omega^2,\nn\\
g_{ab}dx^a dx^b &=& -F(r) dt^2 + B(r) dr^2 + A(r)r^2 d\Omega^2\,.\nn
\eea
Here $q_{ab}$ is an auxiliary metric \cite{Banados:2010ix}, and
we have used the gauge freedom to fix the function in front of the spherical part of the metric $q$. 
We consider perfect-fluid stars with energy density $\rho(r)$ and pressure $P(r)$ such that
\be
T^{ab}\equiv T^{ab}_{\rm perfect\,fluid}=\left[\rho+P\right]u^a\,u^b+g^{ab}P\,,\label{Tmunu_fluid}
\ee
where the fluid four-velocity $u^a=(1/\sqrt{F},0,0,0)$.

We integrate the field equations~(5) and (6) in Ref.~\cite{Banados:2010ix} imposing regularity conditions at the center of the star. 
The series expansion of the field equations at the center of the star contains terms of the form
$\sqrt{(1-\kappa P_c)(1+\kappa \rho_c)}$. Assuming $\rho_c,P_c>0$, $\kappa$ must satisfy two conditions in order to allow for self-gravitating objects:
\begin{eqnarray}
&&P_c\kappa<1\,,\qquad\,\, \text{for $\kappa>0$} \,,\label{lim1}\\
&&\rho_c|\kappa|<1\,,\qquad \text{for $\kappa<0$}\,.\label{lim2}
\end{eqnarray}
Hence, the existence of NSs with $\rho_c\sim8\cdot 10^{17}$~kg\ m$^{-3}$ and $P_c\sim 10^{34}$~N\ m$^{-2}$ strongly constrains the theory, $|\kappa|\lesssim 1\mbox{ m}^5\mbox{kg}^{-1}\mbox{s}^{-2}$.
Furthermore, it is easy to prove that compact objects only exist if $P''(0)<0$. This gives a further constraint depending on $\rho_c$, $P_c$ and $\rho'_c$, whose form is cumbersome, but it is similar to Eq.~\eqref{kcrit}. In particular, the condition is always satisfied for $\kappa>0$.


The field equations are integrated outward up to the radius $R$, defined by the condition $P(R)=0$, where we require the numerical solution to match the exact, and unique, vacuum Schwarzschild solution, $F(r)=B(r)^{-1}=p(r)=h(r)^{-1}=1-2 M/r$, where $M$ is the mass of the star. To match our numerically generated spacetime to a Schwarzschild exterior we use the Darmois-Israel equations~\cite{Israel:1966rt} at the radius, i.e. $[g_{ij}]=0$ and $[K_{ij}(q)]=0$, where $[...]$ is the jump across the surface, $K_{ij}(q)$ is the extrinsic curvature tensor built with the metric $q$, and $i,j=0,2,3$. These matching conditions come from the field equations and the requirement of a well-defined $3-$geometry and give a unique prescription to compute the mass of the spacetime.
%
%
%
%

%
\begin{figure}[htb]
\begin{center}
\begin{tabular}{c}
\epsfig{file=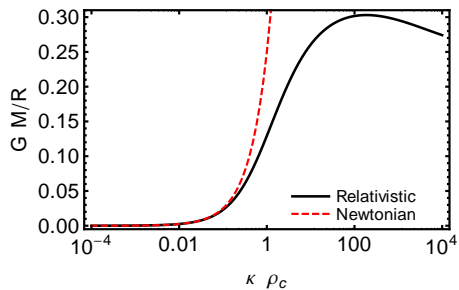,width=6cm,angle=0}
\end{tabular}
\caption{Compactness for pressureless stars in the relativistic theory and in the Newtonian limit, as functions of $\kappa\rho_c$.
\label{fig:p0comp}}
\end{center}
\end{figure}
%
\noindent{\em IIIA1. Relativistic pressureless stars.}
The existence of Newtonian pressureless stars makes it relevant to investigate the existence of similar solutions in the full theory. To this purpose, we set $P\equiv0$. The conservation of the stress-energy tensor simply implies $F(r)=$const. 
The solutions of the field equations then depend only on one parameter, the dimensionless central density $\kappa \rho_c$.

\begin{figure*}[htb]
 \begin{center}
 \begin{tabular}{cc}
 \epsfig{file=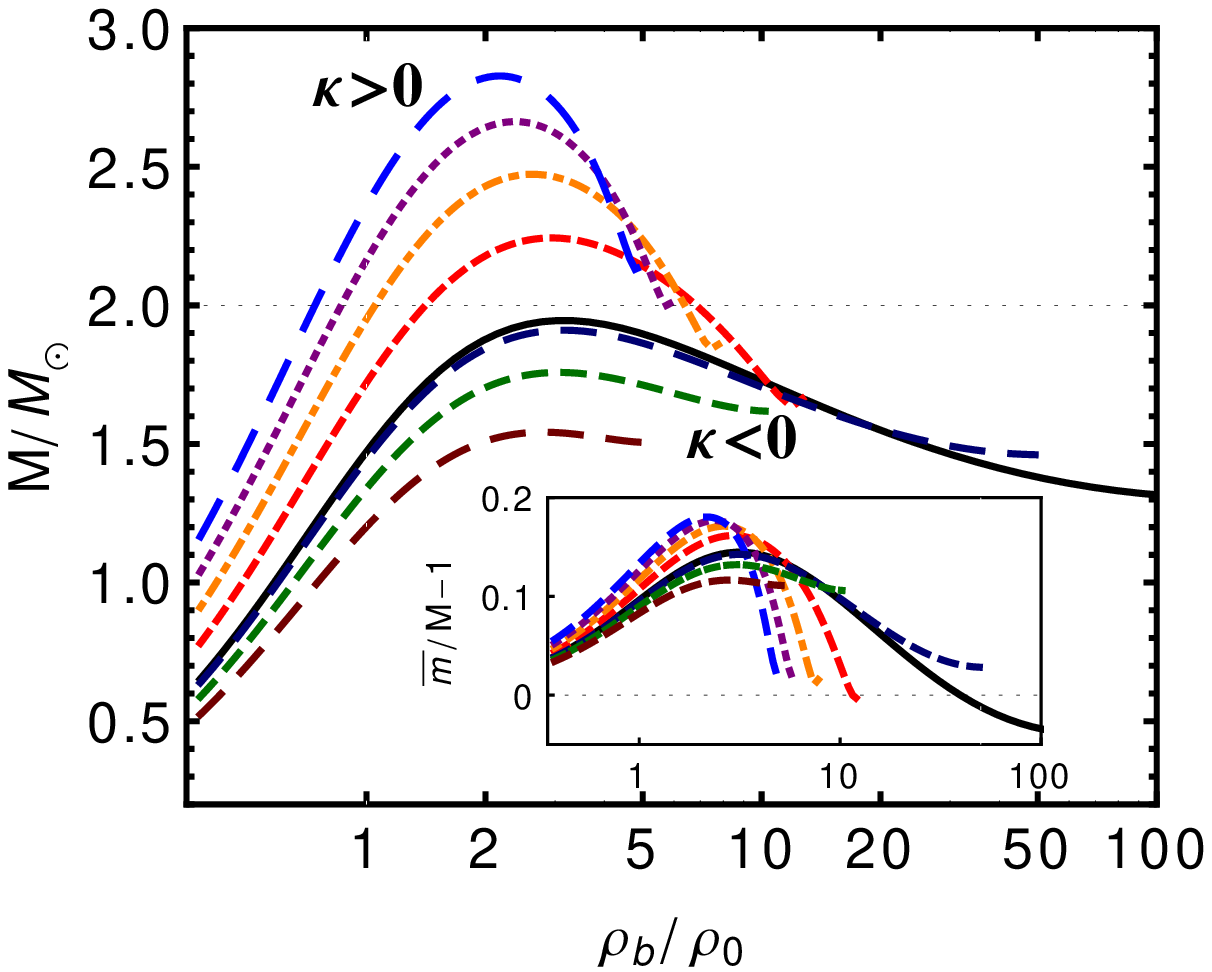,width=6.5cm,angle=0}&\hspace{3cm}
 \epsfig{file=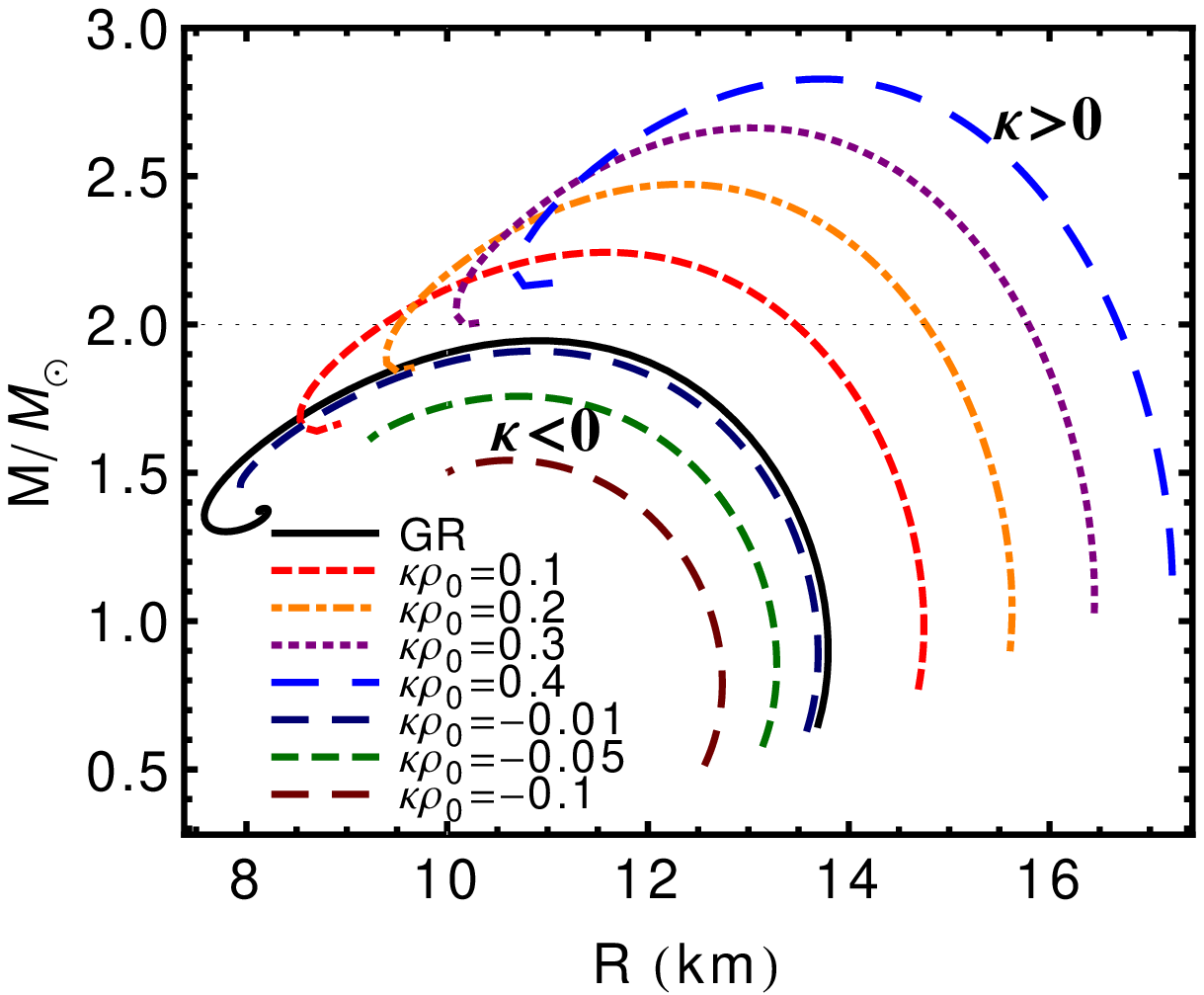,width=6.3cm,angle=0}
 \end{tabular}
 \caption{Polytropic models for different values of $\kappa$. Left panel: mass as a function of the central baryonic density $\rho_b$. Right panel: mass-radius relation. Inset: binding energy as a function of $\rho_b$. Results are normalized by $\rho_0=8\cdot 10^{17}$~kg\ m$^{-3}$, which is a typical central density for NSs. Curves terminate when conditions~\eqref{lim1} or \eqref{lim2} are not fulfilled.
 \label{fig:poly}}
 \end{center}
 \end{figure*}

As shown in Fig.~\ref{fig:p0comp}, for any value of $\kappa>0$, there exists a regular solution which reduces to the Newtonian solution discussed above in the nonrelativistic limit $\kappa\rho_c\ll1$. These solutions have a positive binding energy and can be as compact as $GM/R\sim0.3$ for $\kappa\rho_c\sim200$. Of course they do not exist in GR, while they exist in BF theory because $\kappa>0$ introduces a repulsive gravity contribution. Interestingly, the EOS for dark matter particles is approximately $P\equiv0$. Hence, in this theory self-gravitating objects, purely made by dark matter, can exist and may reach the typical compactness of most compact NSs. 
Furthermore, these objects are stable in the Newtonian limit and it is reasonable to assume that they would remain stable also in the relativistic theory. 
\noindent{\em IIIA2. Polytropic EOS.}
We consider the model 
\be
\rho=n m_b+K\frac{n_0 m_b}{\Gamma-1}\left(\frac{n}{n_0}\right)^\Gamma \,,\quad P=Kn_0
m_b\left(\frac{n}{n_0}\right)^\Gamma \,,\nonumber
\ee
%
with the same polytropic  parameters as in Ref.~\cite{Damour:1993hw}.
Some results are shown in Fig.~\ref{fig:poly} for different values of $\kappa$.
The stellar mass $M$ is shown as a function of the central baryonic density $\rho_b=m_b
n(0)$. In GR, maxima of this curve correspond to marginally stable equilibrium configurations, all
solutions after the first maximum are unstable to radial perturbations (see e.g.~\cite{Shapiro:1983du}). 
This picture may change when $\kappa\neq0$. However, when $\kappa\rho_c\ll1$  our solutions reduce to the non-relativistic ones, for which we proved stability, at least when $\kappa>0$. Hence, we conjecture that properties similar to GR still hold and branches before the first maximum in Fig.~\ref{fig:poly} are likely stable. We leave a detailed analysis for the future.
%

In the inset of Fig.~\ref{fig:poly} we also show the normalized binding energy reads $E_b/M=\bar{m}/M-1$, where $\bar{m}=m_b\int d^3x\sqrt{-g}u^0 n(r)$, is the baryonic mass of the configuration and corresponds to the energy that the system would have if all baryons were
dispersed to infinity. For bound (not necessarily stable) configurations, $E_b>0$.
 %
%
%

Positive values of $\kappa$ tend to enhance the relativistic effects: the maximum mass is larger than in GR and it occurs for smaller central density. Moreover, the binding energy for these models increases with $\kappa$. Negative values of $\kappa$ have the opposite behavior. Remarkably, the most interesting effects show up when $\kappa>0$, i.e. in the same region where singularities seem to be prevented. 

These effects could be observable. Present NS observations constrain the mass-radius relation (e.g.~\cite{Steiner:2010fz}), and
electromagnetic observations of binaries containing X-ray pulsars may in principle constrain the binding energy as well~\cite{Alecian:2003ez}.  The recent discovery of a high-mass NS~\cite{Demorest:2010bx} also rules out many EOS in GR.  
However, these observations could be interpreted in terms of modified gravity at large curvature, rather then invoking exotic EOS in GR.

%
%

\begin{figure}[htb]
\begin{center}
\begin{tabular}{c}
\epsfig{file=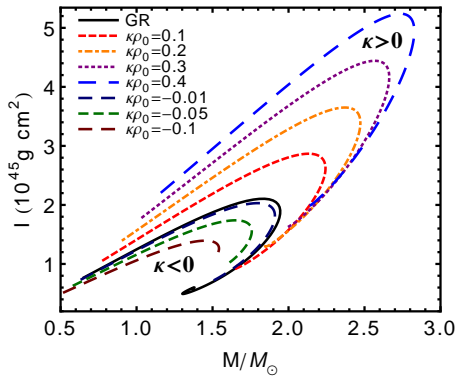,width=6cm,angle=0}
\end{tabular}
\caption{Moment of inertia for polytropic models as a function of the stellar mass for different values of $\kappa$. 
\label{fig:poly_momInertia}}
\end{center}
\end{figure}

\noindent{\bf{\em IIIB. Slowly rotating models.}}
Slowly rotating stars can be constructed from the corresponding static solutions~\cite{Hartle:1967he}. At first order in the rotation, $g_{t\varphi}=-\zeta(r)r^2\sin^2\theta$, $q_{t\varphi}=-\eta(r)r^2\sin^2\theta$ and the stress-energy tensor for a rotating fluid can be built from from Eq.~\eqref{Tmunu_fluid} with 
\be
u^a=\left\{u^t,0,0,\Omega u^t\right\} \,,\quad u^t=\sqrt{-(g_{tt}+2\Omega_{t\varphi}+\Omega^2 g_{\varphi\varphi})}\,,\nn
\ee
where $\Omega$ is the angular velocity of the fluid. The field equations for $\eta$ and $\zeta$ have to be solved by imposing regularity at the center and matching the vacuum solution, $\eta=\zeta=2J/r^3$ at the stellar radius, where $J$ is the angular momentum. In Fig.~\ref{fig:poly_momInertia} we show the moment of inertia $I=J/\Omega$ as a function of the stellar mass.

%
%
%
%
%
%
\noindent{\bf{\em IV. Conclusions.}}
Eddington inspired theories are viable, one-parameter, alternatives to Einstein's gravity. We have shown that in these theories the structure of compact stars is dramatically different from GR, with potentially observable effects. For $\kappa>0$, our results show that BF theory has several remarkable features, e.g. singularities in gravitational collapse may be prevented. The mere existence of compact NSs strongly constrains the theory, $\kappa P_c<1$. Furthermore, in our simple polytropic model, observational determination of the moment of inertia to an accuracy of $10\%$, as it is expected from future observations of the double pulsar~\cite{Lattimer:2004nj}, will place even a stronger constraint, $|\kappa\rho_0|\lesssim0.1$ (cf. Fig.~\ref{fig:poly_momInertia}). We expect that realistic EOS would constrain $\kappa\rho_0$ by the same order of magnitude. 
Remarkably, NSs are the densest matter configurations in the universe, so that these are likely the strongest bounds on the theory.  
Furthermore, it happens that the typical density of a NS, $\rho_0\sim8\cdot 10^{17}~\text{kg }\text{m}^{-3}$, corresponds to the density of the early universe (age $\sim10^{-6}s$), thus the present analysis can put strong constraints on the cosmological effects found in Ref.~\cite{Banados:2010ix}. Several interesting issues, e.g. the collapse in the relativistic theory and black hole formation, the role of realistic EOS, the stability analysis of relativistic stars and possible ergoregion instability of rotating models, are left for future work.

\vspace{0.1cm}
\noindent{\em Acknowledgments.}
  We thank Pedro Gil Ferreira for useful discussions.
  This work was supported by the {\it DyBHo--256667} ERC Starting
  Grant and by FCT - Portugal through PTDC projects FIS/098025/2008,
  FIS/098032/2008, CTE-AST/098034/2008.  

\bibliographystyle{h-physrev4}
\bibliography{Eddington}
\end{document}